\documentclass[10pt]{iopart}\usepackage{iopams}  

\usepackage{graphicx}

\usepackage{mathrsfs} 

\usepackage{epsfig}
\usepackage{eucal} 
\usepackage{amssymb}
\usepackage{amsfonts}
\usepackage{amsbsy}

\begin{document}
\title[]{A Theory of Gravity and General Relativity based on Quantum Electromagnetism}
\author{J.X. Zheng-Johansson}
\address{Institute of Fundamental Physics Research,  611 93 Nyk\"oping, Sweden
}

\def\ai{\a}
\def\Oma{{\Om\hspace{-0.29cm}^{_{\mbox{-}}} \hspace{0.13cm}}}
\def\Omasub{{\Om\hspace{-0.22cm}^{_{\mbox{-}}} \hspace{0.13cm}}}

       \def\Ombar{\Om}
\def\Kbar{K}
\def\fbar{f_d}
\def\Phimbar{\Ccal}

\def\psip{\vphi}

\def\psipp{\xi^\dagsup{}}
\def\psipm{\xi^\ddagsup{}}
\def\citeExp-tests-of-GR{1}
\def\citeEinstein1911{2}
\def\citejxzjied{12}
\def\qvp{q_{_+}}
\def\qvm{q_{_\m}}
\def\qvpm{q_{_\pm}}

\def\ob{{\mathscr{M}}}
\def\obsub{{}}
\def\ka{\kappa}
\def\emptysetft{\mbox{\footnotesize{$\emptyset$}}}
\def\citeExp-tests-of-GR{1}
\def\K{K}
\def\imu{{i\muu}}
\def\mui{{\muu i}}
\def\ip{{j}}
\def\sesub{{\mbox{\tiny{$se$}}}}

\def\musup{{\mbox{\tiny{$\muu$}}}} 
\def\musub{{\mbox{\tiny{$\muu$}}}} 
\def\sumsub{{\mbox{\tiny{$\sum$}}}}

\def\isup{{i}}
\def\mrest{\bar{m}}
\def\oarc{\widetilde}
\def\ine{{\rm{*}}}
\def\ino{{\rm{0}}}
\def\qinst{q^*}
\def\ac{\mbox{{\sf{a}}}}
\def\acb{{\pmb{\ac}}}
\def\phib{\pmb{\phi}}
\def\orl{\overline}

\def\bopsub{{\mbox{\scriptsize{$\bigoplus$}}}} 
\def\Vo{{\ve_\oplus}}
\def\Vol{{V\hspace{-0.29cm}^{_{\mbox{-}}} \hspace{0.13cm}}}
\def\Volp{{V\hspace{-0.29cm}^{_{\mbox{-}}} \hspace{0.13cm}}}
\def\Volsub{{V\hspace{-0.25cm}^{_{\mbox{-}}} \hspace{0.13cm}}}

\def\Vob{\veb_\oplus}
\def\mr{\mathscr{M}}

\def\cmsub{{\rm{cm}}}
\def\dip{1}

\def\j{{}}
\def\jj{\a}
\def\jsub{{\mbox{\tiny{$\j$}}}} 
\def\lamsub{{{}_\mbox{\tiny{$\lam$}}}}

\def\muu{\mu}

\def\phihat{\hat{\phi}}

\def\yhat{\hat{y}}
\def\xhat{\hat{x}}
\def\zhat{\hat{z}}
\def\nhat{\hat{n}}
\def\rhat{\hat{r}}
\def\Rhat{\hat{R}}

\def\nb{\bf{n}}

\def\rcm{r}
\def\rbcm{\rb}
\def\rcmhat{\hat{r}}

\def\yhat{\hat{y}}
\def\vphihat{\hat{\vphi}}

\def\thetahat{\hat{\theta}}
\def\chat{\hat{c}}
\def\thetab{\pmb{\theta}}
\def\vphib{\pmb{\vphi}}

\def\ni{{N}}

\def\rar{\rightarrow}
\def\gtsim{\gtrsim} 
\def\ltsim{\lessim} 

\def\psub{{\mbox{\tiny{$+$}} \hspace{-0.12cm}}}
\def\in{{\mbox{\tiny{$(in)$}}}} 
\def\ellsub{{\mbox{\tiny{$l$}}}} 

\def\prms{{''}}
\def\p{{\mbox{\scriptsize{$+$}} \hspace{-0.03cm}}}
\def\pe{\p e}
\def\pq{\p q}


\def\Ap{\Amp_1}
\def\Asub{{\mbox{\tiny{A}}}}
\def\Msub{{\mbox{\tiny{M}}}}

\def\Npsub{{\mbox{\tiny{$N$}}}}
\def\Nb{n_1}
\def\Np{N}
\def\No{\Ncal_1{}}
\def\Nr{\Ncal}

\def\Nsup{{\mbox{\scriptsize{$N$}}}}

\def\Tsub{{\mbox{\tiny{T}}}}
\def\AAsub{{\mbox{\tiny{A}}}}
\def\Bsub{{\mbox{\tiny{B}}}}
\def\Csub{{\mbox{\tiny{C}}}}
\def\BCsub{{\mbox{\tiny{BC}}}}
\def\Ssub{{\mbox{\tiny{S}}}}
\def\Nsub{{\mbox{\tiny{N}}}}
\def\Osub{{\mbox{\tiny{O}}}}
\def\Psub{{\mbox{\tiny{P}}}}
\def\Esub{{\mbox{\tiny{E}}}}

\def\vsubp{{\mbox{\tiny{v}}}}
\def\vsub{{{}_\v}}

\def\ud{u_1}
\def\udb{\ub_1}

\def\uv{u}
\def\av{u}
\def\sig{\sigma}

\def\pv{{p_{\vsub}}}
\def\vp{\pv}
\def\nv{{n_{\vsub}}}
\def\nvef{{n_{\vsub}^{{\rm ef}}}}
\def\vn{\nv}

\def\rbar{{\bar{r}}}
\def\pbar{{\bar{p}}}
\def\ebar{{\bar{e}}}
\def\abar{{\bar{\a}}}

\def\c{s}
\def\s{i}
\def\qc{q}

\def\Gcal{\mathscr{G}}
\def\Ab{{\bf{A}}}
\def\ab{{\bf{a}}}
\def\gb{{\pmb {\mit g}}}

\def\cb{{\bf c}}
\def\gw{\w}
\def\rr{r}
\def\em{e}
\def\Ocal{\mathcal{O}}
\def\qz{q}
\def\a{\alpha}
\def\abar{{\bar{\a}}}
\def\nee{{\mbox{\scriptsize{-\hspace{-0.3mm}e}}}}
\def\qt{{qt}}
\def\fz{\phi}
\def\ft{\theta}
\def\f{\mathcal{A}}
\def\Acal{\mathcal{A}}

\def\zetab{{\pmb{\zeta}}}

\def\Acalb{\pmb{\Acal}}
\def\Ampb{\Acalb}
\def\Amp{\Acal}

\def\Acalw{\Acal'}

\def\Gscr{\mathscr{G}}

\def\A{A}
\def\vphiq{\psi_q}
\def\etam{{\mbox{\small{${\mathcal{Q}}$}}}}
\def\dash{{\mbox{\tiny{$_-$}}}}
\def\bcal{\mathbin{{\etam}\mkern-8.5mu^{_{\mbox{\small{$\dash$}}}}\hspace{-0.04cm} }}
\def\b{\etam}
\def\rhow{\rho'}
\def\engw{\eng'}
\def\engb{\tilde{\eng}_\vphi}
\def\engqa{\eng_q}
\def\engqan{\tilde{\eng}_{qn}}
\def\Aqa{\tilde{A}_q}
\def\enga{\tilde{\eng}}
\def\velqz{\vel_{qz}}
\def\rhoqz{\rho_{q}}
\def\Dqz{D_{q}}
\def\db{{\bf d}}
\def\pb{{\bf p}}

\def\osub{{\mbox{\small{$0$}}}}
\def\osub{0}

\def\osub{0}
\def\onesub{{{\mbox{\tiny{$1$}}}}}

\def\N{{\mbox{\tiny{$N$}}}}
\def\Nsm{{\mbox{\scriptsize{$N$}}}}
\def\Ncal{\mathscr{N}}
\def\Nscr{\mathscr{N}} 
\def\Fap{\mathscr{F}} 
\def\Fbap{{\pmb{\mathscr{F}}}} 

\def\od{{\rm od}}
\def\eve{{\rm ev}}
\def\vtheta{\vartheta}
\def\phiz{\fz}
\def\vphiz{\phi}
\def\phix{\varphi}

\def\nh{n}
\def\fa{f}
\def\zrm{\mbox{\rm{z}}}
\def\minus{\mbox{-}}
\def\m{{\mbox{-}}}
\def\Rbqt{{\bf R}}
\def\Jb{{\bf J}}
\def\Omb{{\pmb \Om}}
\def\omb{{\pmb \om}}

\def\a{\alpha}
\def\ph{{{\rm ph}}}
\def\ph{{\rm{ph}}}
\def\ub{{\bf{u}}}
\def\v{{\mbox{\footnotesize{\rm{v}}}}}
\def\vmu{{\mbox{\footnotesize{\rm{v}$_\musub$}}}}

\def\vrm{{\rm v}}
\def\vac{\v}
\def\strvac{\supvac}
\def\supvac{\v}
\def\Nsub{{\mbox{\tiny${N}$}}}
\def\Zsub{{\mbox{\tiny${Z}$}}}
\def\kin{{\rm kin}}
\def\Y{Z}
\def\Z{Z}
\def\Pcal{{\mathcal{P}}}
\def\Fcal{{\mathscr{F}}}
\def\Fcalb{{\pmb{\Fcal}}}

\def\bfrak{b_\v}
\def\ii{n}
\def\Esub{{\mbox{\tiny${E}$}}}
\def\cEphi{\Ccal_{\Esub \vphi}}
\def\cEu{\Ccal}
\def\cuphi{A_q}
\def\Ccal{{\mathcal{C}}}

\def\Bcal{{\mathcal{B}}}
\def\Xsub{{\mbox{\tiny${X}$}}}
\def\Ucal{{\mathcal{U}}}
\def\subempty{\empty}
\def\str{{}}

\def\strempt{{}}

\def\prim{\prime}
\def\ArticleLabel-lp{16}
\def\Authorname{J.X. Zheng-Johansson}
\def\Cyc{\zrm_\vphi}
\def\vir{{\rm vir}}
\def\AppA{}
\def\AppChDy{}
\def\AppB{}
\def\AppC{}
\def\citeUnif{xx}
\def\citeHDDMunich{5} 
\def\vel{\upsilon}
\def\ve{\upsilon}
\def\Eb{{\bf{E}}}
\def\Bb{{\bf{B}}}
\def\obs{{\rm obs}}
\def\ev{\epsilon}
\def\ke{\kappa}
\def\Wvel{\mathbin{{\mit\Omega}\mkern-13.mu^{_{\mbox{$-$}}}\hspace{-0.08cm}{}_d }}
\def\wvel{\varpi_d}

\def\q{\mathbin{q\mkern-11mu-}}
\def\empty{{\mbox{\tiny${\emptyset}$}}}

\def\nf{n}
\def\Kcal{{\math{K}}}
\def\Xb{{\bf{X}}}
\def\Pb{{\bf{P}}}

\def\CL{{\mbox{\tiny{{\it C.L}}}}}
\def\imr{{\rm im}}
\def\pl{{\mbox{\tiny{$\|$}}}}
\def\pp{{\mbox{\tiny{$\bot$}}}}
\def\Thm{\vartheta}
\def\TThm{{\mit{\Theta}}}
\def\Thetam{{\mit{\Theta}}}
\def\veli{{\mathscr{V}}}
\def\velib{{\pmb{\veli}}}
\def\Xim{{\mit{\Xi}}}
\def\Xima{\xi}
\def\Vel{W}
\def\Velb{{\bf{\Vel}}}
\def\Pw{{\psitot_p}}
\def\PW{{\Psim_p}}
\def\Mcal{\mfrak}
\def\nablab{{\pmb{\nabla}}}
\def\veb{{\bf{v}}}

\def\mb{m_a}

\def\mP{m_p}

\def\veb{{\pmb{\ve}}}

\def\zb{{\bf{z}}}
\def\xib{{\pmb{\xi}}}
\def\xia{\a}

\def\rb{{\bf{r}}}
\def\yb{{\bf{y}}}
\def\Db{{\bf{D}}}
\def\xb{{\bf{x}}}
\def\d{\delta}
\def\sb{{\bf{s}}}

\def\jb{{\bf{j}}}
\def\fb{{\bf{f}}}
\def\Fb{{\bf{F}}}
\def\wrm{{\rm w}}
\def\med{{\med}}

\def\mfrak{{\mathfrak{M}}}
\def\Mfrak{{\mathfrak{M}}}
\def\Mfk{{\mathfrak{M}}}

\def\Jcal{j}
\def\Jcalb{{\bf{j}}}
\def\Jobs{j_{\rm obs}}
\def\jobs{\Jobs}
\def\jr{j_{\rm obs}}
\def\obs{{\rm{obs}} }
\def\qm{{\rm qm}}
\def\Dscr{D_{\qm}}
\def\om{\omega}




\def\Om{{\mit \Omega}}
\def\g{\gamma{}} 

\def\lamsub{{\mbox{\tiny{$\lam$}}}}

\begin{abstract}

\def\a{\alpha}
\def\Gcal{\mathscr{G}}
\def\lam{\lambda}
\def\Nscr{\mathscr{N}}

Based on first principles solutions in a unified framework of quantum mechanics and 
electromagnetism  we predict the presence of a universal attractive depolarisation radiation (DR) Lorentz force ($F$) between quantum entities, each being either an IED matter particle or light quantum, in a polarisable
dielectric vacuum. Given two quantum entities $i=1,2$ of either kind, of characteristic frequencies $\nu_i^0$, masses $m_i^0=h\nu_i^0 / c^2$ and separated at a distance $r^0$, the solution for $F$ is
$F=- \Gcal m_1^0 m_2^0/ (r^0)^2$,  where $ \Gcal= \chi_\osub^2 e^4/12 \pi^2 \epsilon_0^2 \rho_{_{\lambda}}$; $\chi_\osub$ is the susceptibility and $\rho_{_{\lambda}}$ is the reduced linear mass density of the vacuum. 
This force $F$ resembles in all respects Newton's gravity and is accurate at the weak $F$ limit;
hence $ \Gcal $ equals the gravitational constant $G$. The DR wave fields and hence the gravity are  propagated in the dielectric vacuum at the speed of light $c$; these can not be shielded by matter. 
A test particle $\muu$ of mass $m^0$  therefore interacts gravitationally with 
all of the building particles of a given large mass $M$ at $r^0$ apart,
by a total gravitational force $F= -G M m^0/ (r^0)^2$ and potential $V = -\partial F/\partial r^0 
$. 
For a finite $V$ and hence a total Hamiltonian $H= m^0 c^2 +V$, solution for the eigenvalue equation of $\muu$ presents a red-shift in the eigen frequency $\nu= \nu^0 (1- GM/r^0 c^2)$ and hence in other wave variables. 
The quantum solutions combined with the wave nature of the gravity  further lead to dilated gravito optical distance $r=r^0/(1- GM/r^0 c^2)$ and time  $t=t^0/(1- GM/r^0 c^2)$, and modified Newton's gravity and  Einstein's mass energy relation. Applications of these give predictions of the general relativistic effects manifested in the four classical test experiments of Einstein's general relativity (GR), in direct agreement with the experiments and the predictions given based on  GR.
 \end{abstract}


\def\utot{\mathscr{U}}
\def\Vscr{\mathscr{V}}
\def\V{\mathscr{V}}
\def\Vvq{V_{\v q}}
\def\Vvqo{V_{\v q0}}

\def\uscr{{\mathscr{U}}}

\def\ubscr{{\pmb{\mbox{${\mathscr{U}}$}}}}

      \def\ubscrq{{\mbox{$\pmb{\mathscr{U}}$}\hspace{-0.14cm}}_q}
         \def\ubscrqsq{{\mbox{$\pmb{\mathscr{U}}$}\hspace{-0.14cm}}_q{\hspace{-0.05cm}}}

\def\Zcal{{\mathcal{Z}}}
\def\Kcal{{\mathcal{K}}}
\def\kappab{\pmb{\kappa}}
\def\gd{{\mathcal{G}}}
\def\lb{{\bf l}}
\def\vb{{\bf v}}
\def\Rb{{\bf R}}
\def\pd{\partial}
\def\vphi{\varphi}
\def\vphia{\phi}
\def\vphiab{\pmb{\phi}}
\def\vphiahat{\hat{\phi}}

\def\psitot{{\mathcal{Y}}}
\def\psiR{\widetilde{\psi}}
\def\psiL{\widetilde{\psi}^{{\rm vir}}}
\def\psitotR{\widetilde{\psitot}}
\def\psitotL{\widetilde{\psitot}^{{\rm vir}}}

\def\PhimR{\widetilde{ {\mit \Phi}}}
\def\PsimR{\widetilde{ {\mit \Psi}}}
\def\PsimL{{\widetilde{ {\mit \Psi}}}^{{\rm vir}}}
\def\a{\alpha}
\def\apl{a}
\def\ap{{ap}}
\def\aplF{}
\def\uav{\bar{u}}
\def\D{\Delta}
\def\t{t}
\def\x{x}
\def\z{z}
\def\y{y}

\def\rest{{\rm{rest}}}

\def\th{\theta}
\def\r{{\mbox{\tiny${R}$}}}
\def\re{{\mbox{\tiny${R}$}}}
\def\Fmed{F_{{\rm a.med}}}
\def\med{{\rm med}}
\def\Lw{L_{\varphi}}

\def\Efb{{\bf E}}
\def\Bfb{{\bf B}}
\def\Ac{ \varphi}
\def\Xsub{{\mbox{\tiny${X}$}}}
\def\Xssub{{\mbox{\tiny${X}$}}}
\def\Tssub{{\mbox{\tiny${T}$}}}
\def\Kb{{\bf{K}}}
\def\kb{{\bf{k}}}
\def\Ksub{{\mbox{\tiny${K}$}}}
\def\W{{\mit \Omega}}
\def\Wd{\W_d{}}
\def\Nu{{\cal V}}
\def\Nud{\Nu_d{}}

\def\Eng{\varepsilon}

\def\FT{\Tcal}
\def\Tcal{\tau}
\def\Engvekin{T}

\def\rhoeng{\rho_{_{\Eng}}{}}
\def\eng{\rhoeng{}}

\def\Acuni{\Ac_{{\Ksub}^\dagsup}^{\dagsup}}
\def\unduni{\Ac_{{\Ksub}^\dagger}^{\dagsup}}
\def\Acauni{\Ac_{{\Ksub}^\ddagsup}^{\ddagsup}}
\def\Acunim{{\Ac_{{\Ksub}^\dagsup}^{\dagsup *}}}
\def\undunim{{\Ac_{{\Ksub}^\dagsup}^{\dagsup}}^*}
\def\Acaunim{{\Ac_{{\Ksub}^\ddagsup}^{\ddagsup *}}}
\def\pd{\partial}
\def\Ad{ {\mit \psi}}
\def\psim{ {\mit \psi}}
\def\Kd{K_d{}}
\def\Lam{{\mit \Lambda}}
\def\Lamb{\pmb{{\mit \Lambda}}}

\def\lam{\lambda}
\def\dagsup{{\mbox{\tiny${\dagger}$}}}
\def\ddagsup{{\mbox{\tiny${\ddagger}$}}}
\def\psimKdK{\psim_{\Ksub,\Kdsub}}
\def\w{\omega{}}
\def\wdlow{\omega_d }
\def\g{\gamma{}} 
\def\Phim{{\mit \Phi}}
\def\Psim{{\mit \Psi}}     
\def\Psima{{\mit\Psi}}

\def\arm{{\rm a}}
\def\brm{{\rm b}}
\def\crm{{\rm c}}
\def\drm{{\rm d}}
\def\erm{{\rm e}}
\def\frm{{\rm f}}
\def\grm{{\rm g}}
\def\hrm{{\rm h}}
\def\lf{\left}
\def\rt{\right}
\def\Kdsub{{\mbox{\tiny${K_d}$}}}
\def\psimkd{\psim_{\kdsub}}
\def\psimKd{\psim_{\Kdsub}}
\def\hquad{ \ \ } 
\def\Taum{{\mit \Gamma}}

\section{Introduction}\label{sec-intr}
It has long been established in experiment that Newton's inverse square law of gravity, proposed by I Newton in his {\it Principia} (1686), is accurate for macroscopic objects and in weak gravitational ($g$) field. It has been further established in experiment since around the turn of  the 20th century that Newton's law of gravity is deviated in strong $g$ fields for matter objects and light, whereas light having no rest mass is implicitly subject to  Newton's law only.  The deviations mainly exhibit as general relativistic (g-r) effects manifested in the three or four classical-test experiments \cite{Exp-tests-of-GRa,Exp-tests-of-GRb,Exp-tests-of-GRb2,Exp-tests-of-GRc,Exp-tests-of-GRd}
of  A Einstein's general theory of relativity (GR). Einstein's GR predicts all the phenomena successfully \cite{Einstein1911a,Einstein1911b,Einstein1911c,Exp-tests-of-GRd}.
The cause of gravity yet remains unresolved up to the present, with or without a g-r effect. I. Newton proposed his inverse square law essentially on phenomenological basis. A. Einstein assumed in his GR that a geometric curvature is being produced in the empty-space and time about a material object and is the cause of gravity. GR remains a phenomenological theory. The geometric description of GR is incompatible with the particle and field based description of quantum mechanics (QM),  electromagnetism, and the other three fundamental forces.

QM and GR, along with Newton's gravity in the weak $g$ limit,  have been experimentally corroborated to great extents. The two remain yet un-united \cite{Carlip2001}. 
A unification of the two, under quantum gravity (QG), has been the holly grill of modern physics. Superstring  theory is a much acknowledged contender for QG \cite{Green-etal}. Yet the super-partner particles of the theory remain  to be observed in experiment; the strings and the six or so extra dimensions remain to be demonstrated in experiment, and supplied with physics foundations. As one of other aspects, the tiny strings $\sim 10^{-35}$ m are just points to the particle waves; they do not ultimately resolve the outstanding difficulty related to wave-particle duality of QM. Loop quantum gravity is focused on quantising the space, rather than unifying gravity with the other fundamental forces. QG is desired in black hole and big bang researches.

An Internally Electrodynamic (IED) model of particles, along with a  polarisable vacuuonic dielectric vacuum, has been developed by the author since 2000 using overall experimental observations for particles and vacuum as input information. Based on  first principles solutions for the IED particles  in a unified framework of classical, quantum and relativistic mechanics, it has been possible to  predict a range of the well-established basic properties and relations of particles (see \cite{jxzjied} for a review and the original papers). One of the solutions is a depolarisation radiation (DR) Lorentz force acting between IED matter particles in a polarisable dielectric vacuum \cite{jxzjied-grav}, which 
resembles in all respects Newton's gravity. In this paper, we extend the solution of gravity to including light quanta (Secs \ref{Sec-qmEM}, \ref{Sec-grav-pht}), and the effect of gravity on the dynamical variables of the interacting entities (Sec \ref{Sec-microGR}) within the same unified theoretical framework, which for gravity  essentially consists in
quantum electromagnetism. We thereby obtain a generalised quantum-electromagnetic theory of gravity and general relativity. 
The basic solutions within the  theory
 are then applied to predict the g-r effects manifested in the four classical test experiments of GR (Sec \ref{Sec-exps}). The basic solutions include quantum gravitational waves (composed of  DR fields) from individual matter particles or light quanta; the natural extension of this to an accelerating macroscopic object is a macroscopic gravitational wave, which formal solution we shall describe in a separate paper.

\section{Quantum electromagnetic descriptions of matter particles and light quanta. Dielectric theory of the vacuum}
\label{Sec-qmEM}\label{Sec-grav-pht-1}

We shall work in a three dimensional flat space; this space has a one to one correspondence to a polarisable dielectric vacuum. Let  there be in this space a charge $q$, of charge density $\rho_q$, at  the origin of a  co-ordinate system $x,y,z$ fixed therein.
$q$ is in accelerated motion, forming a current density $\jb_q$, and generates thereby electromagnetic radiation fields $\Eb_\j,\Bb_\j$ according to the Maxwell equations $$\displaylines{
\refstepcounter{equation} \label{eq-Max}
\hfill
\nablab\cdot \Eb = \rho_{q_\j}/\ev_0, \quad
\nablab\times \Bb -(1/c^2) \pd_t \Eb=\mu_0 \jb_q, 
\quad
\nablab \times \Eb+ \pd_t \Bb =0, \quad \nablab \cdot \Bb=0, 
\hfill (\ref{eq-Max})
}$$
where $c (=1/\sqrt{\ev_0 \mu_0}) $ is the speed of  light in the vacuum; $\ev_0$ is the permittivity and $\mu_0$ permeability of the vacuum. In the empty space (\emptysetft\,) left over after removal of the dielectric vacuum, we directly measure the applied radiation  fields $\Eb_{\empty} $ and $\Bb_\empty$. An applied $\Eb_\empty$  at  a location $\rb$ will polarise the vacuum in the vicinity, and induce a vacuum polarisation $\Pb$ and a depolarisation (radiation)  field $\Eb_p$ relative to \emptysetft. Applying the usual dielectric theory \cite{Bottcher:1973,jxzjied-DieVac} here, these are given in terms of the macroscopic fields $\Eb$, $\Bb$ measured in the vacuum as
$$\displaylines{
\refstepcounter{equation} \label{eq-a2p}
\hfill
\Eb_{\empty} = \Eb-\Eb_{p}= \ka_0 \Eb, \quad \Eb_{p} =-\frac{\Pb }{\ev_\empty}=- \chi_{\osub} \Eb,
\quad 
 \chi_\osub=\ka_0-1; 
\hfill 
\cr
\hfill
\Bb_\empty=\Bb-\Bb_p =\ka_0 \Bb, 
\quad
\Bb_p =-\frac{ \Eb_p  \times \cb}{c^2} = -\chi_\osub \Bb. 
\hfill (\ref{eq-a2p})
}$$
$\ka_0(>1)$ is the dielectric constant and $\chi_{\osub} (>0)$ is the susceptibility of the dielectric vacuum against $\emptysetft$; $\ev_0= \ka_0 \ev_\empty$, and $\ev_\empty$ is  the permittivity of the empty space $\emptysetft$. We have implicitly assumed no significant magnetisation of the vacuum by $\Bb_\empty $. $\Bb_p$ is thus a magnetic field purely induced by the time rate of $\Eb_p$ (Maxwell-Ampere's law). In terms of the vacuuonic vacuum model\footnote{
 It can be shown that, combined Lorentz transformations for both the sources and the interferometer arms as of the Michelson-Morley and Kennedy-Thorndike experiments,
render null and constant fringe shifts of light in two-way trips against a substantial vacuum
 (Zheng-Johansson J X 2005   {\it Inference of Basic Laws of Classical, Quantum and Relativistic Mechanics based on  First-Principles of a Minimal Set} 254 pages, unpublished).} \cite{jxzjied,jxzjied-DieVac,jxzjied-VacStru}, 
(a) the vacuum is filled of densely packed, electrically neutral but polarisable entities called vacuuons separated at a mean spacing $b$;   $b \sim 10^{-18} $ m estimated based on the observational shortest wavelength ($1\sim 2 \times 10^{-17}$ m) of electromagnetic radiation; and (b) each vacuuon is composed of a pair of spinning entities called $p$- and $n$- vaculeons of  a charge $+e$ and $-e$, that are located at the core and on a spherical shell about  the core, and are strongly bound electromagnetically.



Assuming  $q$ ($=+e$ or $-e$) is oscillating sinusoidally about $\rb=0$ along $z$ direction, with a frequency $\nu$ and amplitude $\Amp$, Eqs (\ref{eq-Max}) have the usual solutions for $\Eb$ at a distance  $\rb_\j (r_\j,\theta_\j,\vphia_\j)$ sufficiently far from $q_\j$,
which combined with (\ref{eq-a2p}) further yield solutions for $\Eb_p$, given as
 $$\displaylines{
\refstepcounter{equation} \label{eq-a2}
\hfill
\Eb (\rb_\j,t)
=- \frac{a_\j E_{ 0_\j}  \sin \theta_\j  \psip (\rb,t)}{r_\j} \thetahat,
\quad 
E_{ 0_\j}=\frac{ e \Amp_{\j} \om_\j^2  }{ 4 \pi \ev_0 c^2    a_\j    },
\quad
\psip(\rb,t)= \cos (\kb_\j \cdot \rb_\j -\om_\j t + \xia_\osub), 
\hfill
(\ref{eq-a2})
\cr
\refstepcounter{equation} \label{eq-a2pB}
\hfill
 \Eb_p (\rb,t; -\mbox{$\frac{\pi}{2}$})= - \chi_\osub \Eb(\rb,t)
=\frac{  E_{p0}  a \, \sin \theta   \psi_p    }{r} \thetahat, 
\quad  E_{p0} = -\chi_\osub E_0,
  \quad
\label{eq-psig}
\psi_p (\rb,t)
=- \psip (\rb,t; { -\frac{\pi}{2}});  
\hfill 
(\ref{eq-psig})
}$$
and $\Bb  =- \frac{\Eb \times \cb}{c^2} $, $\Bb_p = - \frac{\Eb_p \times \cb}{c^2}$. Here $\om_\j = 2\pi \nu_\j$, $k=\frac{2\pi}{\lam} $, $\lam=c/\nu$,  $\xia_{\osub_\j}$ is an initial phase, and $a_\j$ is a linear dimension defined after (\ref{eq-a1}) below. $q$ may implicitly also be in linear motion, at velocity $\ve$ in $+x$ direction here. The fields ($\vphi$'s) generated oppositely say in $+x,-x$ directions, as $\psip^\dagsup, \psip^\ddagsup$, are thus in general Doppler effected, differently. 
For the DR force in question in Sec \ref{Sec-qmEM}, $\ve$ affects $E_0 $
through $\om$; $\om= \sqrt{\om^\dagsup \om^\ddagsup}= \g \Ombar$, 
where $\g =(1-\ve^2/c^2)^{-1/2}$ and $\Ombar = \lim_{\ve^2/c^2\rar 0} \om$ from  IED solution (see \cite{jxzjied}). $\ve$ does not affect the integration of $\psip^j$'s which are written as $\psip$ in (\ref{eq-a2}c).

The energy density of the $\Eb,\Bb$ fields,  as $\Eb^{(c)},\Bb^{(c)}$ in complex forms, is $\rhoeng=\ev_0 (\Eb^{(c)} \times \Bb^{(c)}) \cdot \cb$. In the dielectric vacuum of an induced elasticity, $\Eb,\Bb$ correspond to a transverse elastic wave, of a transverse (tensile) displacement $\ub_\j =(\Amp_\j  /E_{0\j})\Eb$ and amplitude $\Amp_\j$. So accordingly $\rhoeng=\frac{1}{2} \rho_{\ellsub} \om_\j^2 |{u_\j^{(c)}}|^2$, where $ \rho_\ellsub$ is the  linear (sub-vacuum)  mass density of the perturbed vacuum. Suppose that the $\Eb,\Bb$  comprising one energy quantum $h \nu$ extend through a spherical volume $\Volp$ of radius $r_0$;  $\ell_\j = 2 r_0 =\No_{\j  } \lam_\j  =\Nb b$ contains $\No_{\j} $ wavelengths $\lam$'s and $\Nb$ vacuuon spacings $b$'s. The one energy quantum, or the Hamiltonian of the one quantum, is thus equal to the integration of $\rhoeng$ in $\Vol$, with $d \Vol= r^2 d r d \Oma$, $d \Oma=\sin \theta  d \theta d \vphia$,
$$\displaylines{
\refstepcounter{equation} \label{eq-a1}
\hfill
\Eng_\j= \int_{\Volsub}  \ev_0 (\Eb^{(c)} \times \Bb^{(c)}) \cdot \cb d \Vol   
=\ell \sig_\j \ev_0 E_{0\j}^2
 = \int_{\Volsub}  \frac{1}{2} \rho_{ \ellsub} \om_\j^2 {u_\j^{(c)}}^2d \Vol   
=\frac{1}{2} \lam_\j  \rho_{\lamsub \jsub}  \om_\j^2 \Amp_{\j}^2 
=h  \nu_\j =m_\j c^2.
\hfill
(\ref{eq-a1})
}$$
$\sig_\j = \frac{4}{3} \pi a^2_\j $ represents a  cross sectional  area through which $\Eng$ is transported. 
$\rho_\lamsub 
= \ell \rho_{_{\ellsub}}  /\lam_\j  
 = \No_{\j  } \lam_\j \rho_{_{\ellsub }} /\lam_\j  
=\No  \rho_\ellsub   $ 
is the {\it reduced} linear mass density of the vacuum.

$\Eb,\Bb$ may be the internal components of a simple, single charged matter particle such as a proton or an electron, generated by its charge $q=+e$ or $-e$; $q$ is of zero rest particle mass but is oscillating with a frequency $\nu$ specified by (\ref{eq-a1}) (the IED model). Its $\Eng$ is in general conveyed a fraction $a_q$ by its oscillatory charge, and $a_r$ by its radiation field, with $0\le a_q, a_r \le 1$. So 
$a_q \Eng + a_r \Eng=a_r \Eng=\Eng (=h \nu =m c^2)$, 
given for the extreme case $a_r=1$ and $a_q=0$ in (\ref{eq-a1}), is the total (rest) mass energy and $m$ is the (rest) mass of the resultant IED particle  (at rest). Or, $\Eb,\Bb$ may comprise a light quantum, (originally) emitted by a matter particle of  charge $q=+e$ or $-e$ here. $m$ then represents the dynamical mass, or simply mass, of the light quantum. 
(\ref{eq-a1}) and (\ref{eq-a2}b) give a few further relations to be used:
$$\displaylines{
\refstepcounter{equation} \label{eq-rho}
\hfill
\rho_{\ellsub_\j} =  \frac{e^2 m^2}{6\pi \ev_0 \hbar^2}, 
\quad
\rho_{_\lam}
=  \No_{\j  } \rho_{_{\ellsub \j}}
=  \frac{\No_{\j  } e^2 m^2}{6\pi \ev_0 \hbar^2} 
=\frac{\hbar }{\pi c \Amp^2}. 
\hfill (\ref{eq-rho})
}$$

\section{Newtonian gravity between light quanta and matter particles }
\label{Sec-grav-pht}
We shall in this section derive a general equation of Newton's gravity between a light quantum $\muu$ and an IED matter particle $i$ (Secs \ref{Sec-grav-pht}.1-2) or $j$ (Sec \ref{Sec-grav-pht}.3), their mean mutual gravity (Sec \ref{Sec-grav-pht}.4), and finally between $\muu$ (designating thereof a light quantum or IED matter particle) and an object composed of many IED matter particles (Sec \ref{Sec-grav-pht}.5).  We shall disregard a g-r effect until Sec \ref{Sec-microGR}.

{\it \ref{Sec-grav-pht}.1 Action of  $\muu $ on $i$ \ } 
$\muu$ and $i$ are specified as follows: 
(a) $\muu$ is one of $\Np_\musub$ light quanta that are being constantly (re-) generated by an array of point sources in the vicinity of the origin  on the $yz$ plane, at $x=0$, in the co-ordinate system $x,y,z$. 
Its source has an oscillation frequency $\nu_\musub$ and amplitude $\Ampb_\musub$. Let specifically the sources be pairs of  dipole charges $q_\musub=\pm e$ of the vacuuons comprising the vacuum here,  referred to as  {\it virtual } sources. They are induced by the plane wave $\Eb_\musub$ fields propagated to  $x=0$ at time $t=0$,  that were generated say at $-x_1$ at time $-t_1$ earlier 
by some natural sources of unit charges $\pm e$.
(b) $i$ is one of $\Np_i$ IED matter particles located at a mean position $x =r$ on the $x$ axis. $i$ has a mass $m_i$, and charge $q_i=+e$ or $-e$ oscillating with a frequency $\nu_i= m_i c^2/ h$ (Eq \ref{eq-a1}) and amplitude $\Ampb_i$. 
(c) $\Ampb_\musub$ and $\Ampb_i$ are along the $z$ direction during a brief time under immediate consideration. So both sources $\eta=i$ and $\muu$ generate fields $\Eb_\eta,\Bb_\eta$ and $\Eb_{p_\eta},\Bb_{p_\eta}$ according to  (\ref{eq-a2})-(\ref{eq-a2pB}) in their local co-ordinate systems,  $x,y,z$ for $\muu$, and a parallel set of axes $x_i,y_i,z_i$   with an origin fixed at  $x=r$ for $i$. $\Eb_\musub,\Bb_\musub$ ($\Eb_i,\Bb_i$) are assumed shielded, not reaching $i$ ($\muu$). And, 
(d) $i$ and $\muu$ are in dynamical equilibrium.

The fields $\Eb_{p_\musup} =\Eb_\musub-\Eb_{\empty_\muu}$ and $  \Bb_{p_\musup}=\Bb_\musub-\Bb_{\empty_\muu}$  (Eq \ref{eq-a2pB}) generated by  $q_\musub$ at $x=0$ at time $t=0$  will propagate to  $q_i$  at $x=r$ at time $t$, and  act on $q_i$  a (momentary) Lorentz force $ \Fb_{tot_{\mui}} =q_i \Eb_{p_\muu }  + \Fb_{\mui} $. $q_i \Eb_{p_\muu }  $ drives  $q_i$  into a transverse  motion at  velocity $ \veb_{p_i} $ given as
$$\displaylines{
\refstepcounter{equation} 
\label{eq-ves}
\hfill
\veb_{p_i}
= \veb_{i}- \veb_{\empty_i} 
=  \frac{ \tau_\muu q_i  \Eb'_{p_\muu} }{2 \pi m_i}, 
\quad
\veb_i
=  \int \frac{q_i \Eb_{\musup } }{m_i}  d t 
=  \frac{ \tau_\muu q_i  \Eb'_\muu  }{2 \pi m_i}, \quad
\veb_{\empty_i}=  \frac{ \tau_\muu q_i  \Eb'_{\empty_\muu} }{2 \pi m_i}, 
\hfill (\ref{eq-ves})
}$$
where $\tau_\musub=1/\nu_\musub$, $\Eb'_\muu (\rb,t) =\frac{ a_\muu E_{0_\muu} \sin \theta }{r}  \cos (\xia_\musub +\frac{\pi}{2}) \zhat \equiv \Eb_\muu (\rb,t; +\frac{\pi}{2})$ given after (\ref{eq-a2}a), etc.;  $\xia_\musub=\kb_\muu \cdot \rb -\om_\muu t  $; we have set  $\a_{0_\musub}=0$ for the initial phase.
$$\Fb_{\mui}  
=q_i  (\veb_i \times \Bb_{\musup}  -  \veb_{ \empty_i }  \times \Bb_{\empty_\musup}) 
=q_i( \frac{ \tau_\muu q_i    }{2 \pi m_i})  [ \Eb'_{\musub} \times \Bb_\musub - (\Eb'_{\musub} -\Eb_{p_\muu}'   )\times (\Bb_{\musup} -\Bb_{p_\muu}) ]
$$
is the difference between a Lorentz magnetic force acted by $\Bb_\muu$ in the vacuum, and one by $\Bb_{\empty_\musup}$ in \emptysetft, which we know to each obey the right hand rule. Setting $\Eb'_\musub=0$, $\Bb_\musub=0$, we obtain the final Lorentz magnetic force acted by $\Eb_{p_\musub}', \Bb_{p_\musub}$  on $q_i$, referred to as DR (Lorentz) force,
$$\displaylines{
\refstepcounter{equation} 
\label{eq-a3.1}
\hfill
 \Fb_{\mui}  
= -q_i \veb_{p_i} \times \Bb_{p_\muu} 
= -\frac{\tau_\muu q_i^2 \Eb'_{p_\muu}\times \Bb_{p_\muu} }{2\pi m_i} 
= -\frac{ \tau_\muu q_i^2 \chi_\osub^2 \Eb'_\musup \times \Bb_\musup }{2\pi m_i} 
=- \frac{\Kcal_\mui f(\xia_\musub)}{r^2} \xhat, 
\hfill
\cr
\hfill
\Kcal_\mui
= \frac{\tau_\muu  q_i^2 \chi_\osub^2  a_\muu^2 E^2_{0_\muu}  \sin^2\theta}{2\pi  m_i c  } 
= \frac{ \tau_\muu  q_i^2 \chi_\osub^2 a_\muu^2 \lf(\frac{q_\muu \Amp_{\muu} \om_\muu^2  
}{ 4 \pi \ev_0 c^2    a_\muu    }\rt)^2 
\cdot (\frac{1}{2} \lam_\muu \rho_{_{\lam_\muu} } h^2)
 }{2\pi  m_i c  \cdot (\frac{1}{2} \lam_\muu \rho_{_{\lam_\muu}} h^2)
} 
= \frac{ \chi_\osub^2 e^4   m_\muu^3  }{4\pi \ev_0^2 h^2 \rho_{_{\lam_\muu}}  m_i    }, 
\hfill (\ref{eq-a3.1})
}$$
 $f(\xia_\musub)= |\sin( \xia_\musub)   \cos(\xia_\musub)| $.
(\ref{eq-ves}a) for $\veb_{p_i} $, (\ref{eq-a2}b) for $E_{0_\muu}$, $\theta=\frac{\pi}{2}$
 ($i$ is essentially a point on the $x$ axis), $q_i =q_\muu = \pm e$, and the identity relations (\ref{eq-a1}) have been used. Based on (\ref{eq-a3.1}), $ \Fb_{\mui} $ is pointed from $i$ to $\muu$, hence an attraction, irrespective of the signs of $q_i, q_\muu$; and $\Fb_{\mui}$ in relation to $\veb_{p_i}$ and $\Bb_{p_\muu}$ obeys a {\it left hand rule}.

For a large $\Np_\musub$ number of $\muu$'s being constantly (re-)generated at $x=0$ over a long time $\D t (>>\No_{\musub} /\nu_\musub, \No_{i} /\nu_i $), firstly  the $\xia_{\osub_\musub}$'s of different $\muu$'s are independent random variables. These reduce their relative radiation intensities by a factor $f_0=\cos^2 \a_{0_\musub}$, rather than producing coherent waves. Secondly, under the influence of random environmental fields, the amplitude of each original natural source, $\Ampb_\musub{}_1$ at $-x_1$,  hence $\Amp_\musub = \frac{a}{|x_1|} |\Ampb_\musub{}_1|$ at $x=0$, would explore in $\D t$ all possible orientations   $\Oma(\theta, \vphia)$. Its projection in the $yz$ plane,  becoming $\Amp_{\musub_{yz}}=  \Amp_\musub (\sin^2 \theta \sin ^2 \vphia + \cos^2 \theta)^{1/2} $ at $x=0$, is now responsible for producing  a DR force  given otherwise by (\ref{eq-a3.1}).  Including the two features,  $F_{\mui} $ is written as 
$F_{\mui (\Npsub_\musub)} 
=- (\Amp_{\musub_{yz} }^2 /\Amp_\musub^2) \Kcal_\mui f_0 f (\xia_\musub) /r^2$. 
Its average over $\Oma$ in (0,$4\pi$), and $\xia_{\osub_\musub},\xia_{\musub}$ in $(0,\pi)$ is
$$\displaylines{
\refstepcounter{equation} 
\label{eq-a3} \label{eq-Fm-mui}
\hfill
\langle  F_{\mui }\rangle
= \frac{1}{\pi^2} \int_{0}^\pi  \int_{0}^\pi(\frac{1}{4\pi} \int_{\Omasub} F_{\mui (\Npsub_\musub)}d \Oma) d \xia_{\osub_\musub} d \xia_{\musub}
= - \frac{\Kcal_\mui \Pcal_{yz} \bar{f}_{0}\bar{f} }{ r^2} 
= -  \frac{  \Kcal_\mui}{3\pi r^2}, 
\hfill (\ref{eq-Fm-mui})
}$$
where $\Pcal_{yz} 
=\frac{1}{4\pi}\int_\Omasub (\Amp^2_{\musub_{yz}}/\Amp_\musub^2) d \Oma
= \frac{2}{3}$, 
$\bar{f}_{0} =  \frac{1}{\pi}\int_{0}^{\pi} f_0 d \xia_{\osub_\musub}=  \frac{1}{\pi}\int_{0}^{\pi} \cos^2 \xia_{\osub_\musub} d \xia_{\osub_\musub}=\frac{1}{2}$, 
and 
$\bar{f} 
= \frac{1}{\pi}\int_{0}^{\pi} f(\xia_{\musub})d \xia_{\musub}
= \frac{1}{\pi}\int_{0}^{\pi} |\sin \xia_{\musub}  \cos \xia_{\musub} |d \xia_{\musub}
=\frac{1}{\pi} $. 

{\it \ref{Sec-grav-pht}.2  Action of $i$ on $\muu$ \ }
The vacuum about $x=0$ is electrically and dynamically characterised by the $\muu$-disturbance as follows:
(i) $\Eb_\musub$ polarises the vacuuons in the volume $ \Vol_\musub =\Nb b \sig  =\Nb b \cdot (n_y b) \cdot (n_z b) $ occupied by the $\muu$ (wave) train. We find  $\Nb  n_y $ polarised vacuuons, and hence $\Nb  n_y $ units of polarisation charges $ +e$'s (or $-e$'s)  on each enclosing Gauss surface perpendicular to $\Eb_\musub$ along the $z$ axis. And (ii) the $\Nb  n_y $ effective oscillators are driven  by  $\Eb_\musub$ into oscillations at the frequency $\nu_\musub$ of $\Eb_\musub$, and hence imparted with a dynamical mass $\D m_\musub = \frac{m_\musub}{ \Nb n_y }  $ each; $m_\musub=h\nu/c^2$ (Eq \ref{eq-a1}).   $(\Nb n_y e) \times \D m_\musub =e m_\musub (\propto \sqrt{\rho_{\lamsub_\musub}})$ returns correctly the $e$ and $m_\musub$ of the original natural source.  The last identity relation renders an equivalent  single vacuuon (virtual source) $\v_\musub$  representation to be used:  $\v_\musub$ is instantaneously located at $x=0$; it carries all the mass $m_\musub$ and polarisation (dipole) charges $(q_\musub=)q_+,q_\m=+e, -e$, oriented along $\Eb_\musub$ direction. 

$\Eb_{p_i}, \Bb_{p_i}$ generated at $x_i=0$ ($x=r$) and $t=0$, will propagate to $\muu$ at $x_i=-r$ ($x=0$) at time $t$, acting on either charge $q_{+}$ or $q_{\m}$ of $\v_\musub$ a Lorentz force $ \Fb_{tot, \imu \pm }=\qvpm \frac{1}{2} \Eb_{p_i}+  \Fb_{\imu_\pm } $. $\qvpm \frac{1}{2} \Eb_{p_i}$ drives $q_{\pm}$ into motion 
according to $m_\musub d\veb_{p_{\muu \pm}}/d t = \pm e\frac{1}{2} \Eb_{p_i}$. The integrated velocities are $\veb_{p_{\muu \pm}} =  \int \frac{\qvpm\frac{1}{2} \Eb_{p_i}}{ m_\muu } dt = \frac{\tau_i \qvpm \frac{1}{2} \Eb_{p_i}' }{2\pi m_\muu}$, where $\tau_i=1/\nu_i$, $\Eb'_{p_i}(0,t)= \Eb_{p_i}(0,t;\frac{\pi}{2})$. $\Fb_{\imu_\pm} $ is a  DR Lorentz force acted by $i$ on  $q_{_+}$ or $q_{_\m}$ of   $\v_\musub$. This force on both charges, hence on $\v_\musub$, is given as,  applying directly the left hand rule (Sec \ref{Sec-grav-pht}.1),
$$\displaylines{
\refstepcounter{equation} \label{eq-Fimu-1}
\hfill
\Fb_{\imu }=\Fb_{\imu_+}+\Fb_{\imu_\m}
=-(\qvp  \veb_{p_{\muu_+ }} +\qvm \veb_{p_{\muu_\m }}) \times \Bb_{p_i}
=-\frac{\tau_i ( \qvp^2+\qvm ^2)  
\frac{1}{2}\Eb_{p_i}'  \times \Bb_{p_i}   }{2\pi m_\muu} 
= \frac{\Kcal_\imu f(\xia_i )}{r^2} \xhat,
\hfill 
\cr
\hfill
\Kcal_\imu 
= \frac{\tau_i \chi_\osub^2 q_\musub^2 a_i^2 E_{0_i}^2 \sin ^2 \theta_i  }{ 2\pi m_\muu c}
=\frac{  \tau_i \chi_\osub^2 e^2   a_i^2 
(\frac{e  \Amp_i  \om_i^2
}{4\pi\ev_0 c^2 a_i} )^2 
\times 
\frac{1}{2} \lam_i \rho_{\lamsub_i} h^2
}{ 2\pi m_\muu c  
\times \frac{1}{2} \lam_i \rho_{\lamsub_i} h^2      } 
=\frac{\chi_\osub^2 e^4 m_i^3}{
4 \pi \ev_0^2 h^2  \rho_{_{\lamsub_i  }}   m_\muu   },
\hfill (\ref{eq-Fimu-1})
}$$
and $f(\xia_i )= |\sin( \xia_i) \cos(\xia_i) | $, where $\xia_i = \kb_i \cdot \rb_i -2\pi \nu_i t$ given for $\a_{0_i}=0$. And  $r_i=r+\d x  \dot{=}r$ in the denominator for  $r >> \d x =c \No_i/\nu_i$, (\ref{eq-a2}b) for $E_{0_i}$  $\theta_i=\frac{\pi}{2}$, and (\ref{eq-a1}) for $ m_i,m_\musub$ have been used. For the $\Np_i$ IED particles measured over long time $\D t$, with $\Ampb_i$ and $\xia_{0_i}$ being random variables similarly as of $\muu$, the average DR Lorentz force acted by $i$ on $\muu$ is similarly given as
$$\displaylines{
\refstepcounter{equation} \label{eq-Fimu-av}
\hfill
\langle  F_{\imu} \rangle  
=\frac{1}{\pi^2}\int^\pi_0 \int^\pi_0 (\frac{1}{4\pi} \int_{\Omasub} 
 \frac{\Amp_{i_{yz}}^2}{\Amp_i^2} f_0 F_{\imu} d \Oma) d \xia_{\osub_i} d \xia_{i} 
=  \frac{\Kcal_{\imu} \Pcal_{yz} \bar{f}_0 \bar{f} }{ r^2}  
=  \frac{ \Kcal_\imu}{3\pi  r^2}. 
\hfill
(\ref{eq-Fimu-av})
}$$

{\it \ref{Sec-grav-pht}.3 Action of $\muu$ on $\ip$ \  }
$j$ is identical to $i$ but is located at $y=r$ on the $y$ axis perpendicular to the $\muu$  path along the $x$ axis. Each point on the $\muu$ train (of length $\ell_\musub$) serves as a new virtual source re-generating (spherical) radiation fields; a point $x$ on it (say in $[-\frac{1}{2}\ell_\musub, \frac{1}{2}\ell_\musub]$) is at a distance $\rb' = x \xhat  +r \yhat + \d z \zhat $ to $\ip$. Here the $\Eb_{p_\musub}' $ and $\Bb_{p_\musub} $ along the $z$ and $x$ directions (the $\psi_{p_\musub}$ waves) regenerated at time $t=0$ will propagate to $j$ at $y=r $, across  a distance $\rb'$ at time $t' = |\rb'|/c$,  and act on $j$ a DR Lorentz force given similar to  (\ref{eq-a3.1}) as, 
with  $m_j=m_i$, $q_j^2=q_i^2=e^2$,
$$\displaylines{
\refstepcounter{equation} 
\label{eq-Fbmuj1}
\hfill
\Fb_{\musub j}
= - q_j \veb_{p_j}   \times \Bb_{p_\musub} 
= -\frac{\tau_\musub q_j^2 (-E'_{p_\musub} \zhat) \times (-E_{p_\musub}  \xhat)}{2\pi m_j c}
\dot{=}-\frac{\Kcal_{ui} f(\xia_\musub')}{r^2}  \yhat,
\hfill
(\ref{eq-Fbmuj1})
}$$ 
where $\veb_{p_j} = \frac{\tau_\musub q_j \Eb_{p_\musub}' }{2\pi m_j} $, $\Eb_{p_\musub}' =-E'_{p_\musub} \zhat  = -\frac{a E_{0_\musub}   \sin \a_\musub'   }{r'}  \zhat $, $\Bb_{p_\musub } = -\frac{E_{p_\musub} }{c} \xhat $ $= -\frac{a E_{0_\musub}  \cos \a_\musub' }{r'} \xhat  $; $ \Kcal_{\musub i}$  is as given by (\ref{eq-a3.1}); $f(\a_\musub') =$ $ | \sin \a_\musub' $ $ \cos \a_\musub' |$, $\a_\musub' = \kb_\musub \cdot \rb' - \om_\musub t' + \a_{0_\musub}' $, and $\a_{0_\musub}'=k_\musub x - \om_\musub\cdot 0 +\a_{0_\musub}$. The last of Eqs (\ref{eq-Fbmuj1}) is given for $r>> \ell_\musub, \d z$, so $\rb'\dot{=}r \yhat $ and all the $\psi_{p_\musub}$ wave fields from different points $x$ on each  $\muu$ train ($\muu$ quantum) will arrive at $\ip$ at $y=r$ at essentially the same time $t=r/c$. $\a_{0_\musub}'$ is thus inconsequential and may be set to $\a_{0_\musub}'=0$.

For the $\Eb_\musub, \Bb_\musub $ originally  emitted from  a (large) $\Np_\musub$ natural sources (at $-x_1$), the projection of $\Ampb_\musub$, $\Ampb_{\musub_{yz}}$, hence  $\Eb_\muu $   in the $yz$ plane  about $x =0$ here is randomly oriented at an angle $\vtheta $ (and $\Bb_\musub$ at $\frac{\pi}{2}-\vtheta$) to the $z $ axis; and so are  $\Eb_{\musub}' $ and   $\Eb_{p_\muu}' $.  The $z$-projection of the $\Eb'_{p_\musub} $ $ \Eb'_{p_\muu, z} = -E_{p_\muu}' \cos \vtheta  \,  \zhat $, and the induction magnetic field along the $x$ direction, $\Bb_{p_\muu, x }= - \frac{|\Eb_{p_\muu, z}|}{c} \, \xhat  =  \frac{|\Eb_{p_\muu}| \cos \vtheta}{c} \xhat  $, are directly responsible for producing a DR force on $j$ at time $t$:
$$\displaylines{
\refstepcounter{equation} 
\label{eq-Fmui-p2}
\hfill 
\Fb_{\musub \ip (\Npsub_\musub)}
 = -q_{\ip} \veb_{p_{\ip}, z}  \times \Bb_{p_\muu, x} 
=-\frac{\Amp_{\musub _{yz} } ^2  }{ \Amp_\musub^2}  |\Fb_{\musub j}|
   \cos^2 \vtheta\, \yhat 
=- \frac{    (\Amp_{\musub _{yz} }^2/\Amp_\musub^2 )  \Kcal_{\mui}   f(\a_\musub') f_\vtheta 
   }{r^2} \, \yhat,
\hfill (\ref{eq-Fmui-p2})
}$$ 
where $f_\vtheta = \cos^2 \vtheta$. For a large $\Np_\musub$ plane-waves propagated to  $x=0$ at time $t=0$, $\vtheta $ may assume all possible values in $(0, 2 \pi)$. The average of $F_{\musub \ip}$ over $\Np_\musub$ quanta over long time $\D t$ is therefore 
$$\displaylines{
\refstepcounter{equation} 
\label{eq-Fmui-p3}
\hfill 
\langle F_{\musub \ip } \rangle
= - \frac{1}{\pi^2} \int_{0}^{\pi}  \int_{0}^{\pi}  \frac{1}{4\pi} \int_\Omasub 
F_{\musub \ip (\Npsub_\musub) }
d \Oma d \xia'_{\musub} d \vtheta 
=-\frac{\Kcal_\mui \Pcal_{yz} \bar{f} \bar{f}_\vtheta }{ r^2}
=-\frac{ \Kcal_\mui}{3\pi r^2}, 
\hfill (\ref{eq-Fmui-p3})
}$$
where $\bar{f}=\frac{1}{\pi}$ as before; $\bar{f}_\vtheta=\frac{1}{\pi}\int^\pi_0 \cos^2 \vtheta d  \vtheta=\frac{1}{2}$.

From (\ref{eq-Fmui-p3}) and (\ref{eq-Fm-mui})  we have $\langle F_{\muu \ip} \rangle =$ $\langle F_{\muu i} \rangle $. The same features underling the actions $j$ on $\muu$ and 
$i$ on $\muu$ inevitably yield $\langle F_{j \musub } \rangle =$ $\langle F_{i \muu } \rangle $. We shall hereafter refer to the $\muu$-$i$ interaction only.  If $\muu$ is also an IED particle, we could have obtained $\langle F_{\imu }\rangle$, say,  directly  from interchanging the subscripts of $\langle F_{\mui}\rangle$, to be given exactly as  (\ref{eq-Fimu-av}). Unless  otherwise specified we shall hereafter make no distinction between a light quantum and (IED) matter particle.

{\it \ref{Sec-grav-pht}.4 Average mutual DR Lorentz force  \ }
$\langle F_{ \mui} \rangle$ and $\langle F_{\imu} \rangle$ are (attractive) action and reaction forces between $\muu$ and $i$. Under condition (d), the two forces must  be equal in amplitude (and opposite in direction), and hence in turn equal to their mean,
$$\displaylines{
\refstepcounter{equation} \label{eq-Fm-mui-av}
 \label{eq-Fm-mui-av2}\label{eq-G} 
\hfill
F = - \sqrt{ |\langle F_{\mui} \rangle|| \langle F_{\imu} \rangle| }
=- \frac{\sqrt{\Kcal_{\mui} \Kcal_{\imu} }}{3\pi r^2}
=-  \frac{ \Gcal m_\muu m_i }{r^2},
\quad
\Gcal =  \frac{  \chi_\osub^2 e^4  }{12 \pi^2 \ev_0^2  h^2 \rho_{_\lam} },
\hfill (\ref{eq-Fm-mui-av})
\cr
\refstepcounter{equation} \label{eq-G2} 
\hfill
\rho_{\lamsub} 
=\sqrt{\rho_{\lamsub_i }  \rho_{\lamsub_\muu}  }
= \frac{\No e^2 m_i m_\musub }{6 \pi \ev_0 \hbar^2},
\quad
\chi_\osub^2 
=\frac{8\pi^3 \ev_0 \No \Gcal  m_i m_\musub}{e^2},    
\quad
\frac{\chi_\osub^2}{\rho_\lamsub}
= \frac{12 \pi^2 \ev_0^2 h^2 \Gcal}{e^4},
\hfill  (\ref{eq-G2})
}$$
and $\No=\sqrt{ \No_{i}\No_{\muu} }$; $\rho_{\lamsub_i}, \rho_{\lamsub_\musub}$ are as given by (\ref{eq-rho}b). The negative sign indicates that $F $ is  an attraction, as $\langle F_{\mui} \rangle$ and $  \langle F_{\imu} \rangle$ are. $F$ given by the inverse square formula (\ref{eq-Fm-mui-av}a) resembles directly Newton's gravitational force, acting between two quantum entities (a light quantum or IED matter particle each) $\muu$, $i$ here; hence $\Gcal$ corresponds to the gravitational constant $G$, $\Gcal=G$.
Based on (\ref{eq-psig}), $\Eb_p, \Bb_p$ are propagated at the speed of light $c$, and so is the gravity $F$.

{\it \ref{Sec-grav-pht}.5 Action on $\muu$ by a macroscopic object $M$  \ }
$M$ is composed of large $\Np_i$ number of IED matter particles $i$'s each as specified by  (a)-(d), Sec \ref{Sec-grav-pht}.1; it has a mass   $M= \sum_{i=1}^{\Np_i} m_i$. $i$ are atomic electrons and protons, and (for the unit charges used here)  electrons and protons comprising the atomic neutrons. Assume as in typical applications that $M$ is uniform and spherical, of a radius $R$; its mass centre is at $\rb$. $\muu$ is a light quantum or IED matter particle located at $\rb=0$. $\Eb_{p_\musup}$, $\Bb_{p_\musup}$ of $\muu$ are the de-polarisation and induction magnetic fields produced internal of (the polarised vacuuons comprising) the dielectric vacuum, as contrasted to density fluctuations of the vacuum; assume no work has been down by the resultant $F$. As such, $\Eb_{p_\musup}$, $\Bb_{p_\musup}$ can not be absorbed, hence nor be shielded by the matter particles on their path. Each particle $i$ of $M$ therefore sees directly $\muu$, and {\it vice versa}. The total DR Lorentz force between $\muu$ and  $M$ is thus 
$$\displaylines{
\refstepcounter{equation} \label{eq-Fg2}
\hfill
F=- \sum_i \sqrt{|\langle F_{\mui } \rangle ||\langle F_{\imu } \rangle |} \, \frac{\rb_\mui}{r_\mui}
= -\Gcal  |\sum_i \frac{ m_i \rb_{\muu i}}{r_{\muu i}^3}| m_\muu 
= -\frac{\Gcal M m_\muu }{ r^2}.  
\hfill (\ref{eq-Fg2})
}$$
$\Gcal $ for each $\muu,i$ pair involves  $\rho_{\lamsub(\musub i)}$, $\chi_{\osub (\musub i)}^2$ (Eq \ref{eq-Fm-mui-av2}b) which  are separately dependent of  $m_\musub, m_i$ (Eqs \ref{eq-G2}a,b), hence indicated by  ${}_{( \muu i)}$ here. Based on (\ref{eq-G2}c), the  ratio $ \chi_{\osub( \musub i)}^2/\rho_{\lamsub (\musub i) }$  is independent of the masses, and so should be that of  $\chi_\osub^2$, $\rho_\lamsub$ for $M$ and $\muu$:
$ 
\frac{\chi_{\osub (\musub 1)}^2}{\rho_{\lamsub (\musub 1) }}
=\ldots
=\frac{\chi_{\osub}^2}{\rho_{\lamsub  }}
 = \frac{12 \pi^2 \ev_0^2 h^2 \Gcal}{e^4}.
$
For the vacuum occupied by $M$, $\rho_\lamsub$ say satisfying the equalities above may be expressed by
 $$\displaylines{
\refstepcounter{equation} \label{eq-Fm-muM-rho-chi02}
\hfill
\rho_\lamsub = \frac{1}{\Np_i}\sum_{i=1}^{\Np_i} \rho_{\lamsub (\musub i)}
= \frac{\sqrt{\No_{\musub} }  \langle \sqrt{\No_{i}}  \rangle   e^2  M m_\muu }{6\pi \ev_0 \hbar^2 \Np_i}, 
\quad
\langle \sqrt{\No_{i}} \rangle  =\sum_{i=1}^{\Np_i}   \sqrt{\No_{i}} \frac{m_i}{M}.   
\hfill (\ref{eq-Fm-muM-rho-chi02})
}$$

\section{Generalised theory of gravity and relativity}
\label{Sec-microGR}
We shall in this section generalise the theory of gravity of Sec \ref{Sec-grav-pht}  by including the effect of gravity on the dynamical variables of a test particle $\muu$  (Secs \ref{Sec-microGR}.1-2) and test macroscopic object (Sec \ref{Sec-microGR}.3). We refer to this effect as the {\it general relativistic } (g-r) effect in this work. The g-r effect manifestly coincides with the additional content of Einstein's GR over Newton's gravity; the usual light quantum and the IED particle  employed in Sec \ref{Sec-grav-pht} are already intrinsically special relativistic and governed by the Lorentz transformations. To facilitate the discussion, we re-locate $M$ at $r=0$ and $\muu$ at  a distance $\rb$ from it in the co-ordinate system $x,y,z$. And we re-express Newton's gravity  here using  "proper" dynamical variables $r^0, m^0$ measured at the limit $g \cdot r^0 \rar 0$, indicated by  a superscript $0$, acting along  $\rb$ direction as,
$$\displaylines{
\refstepcounter{equation} \label{eq-F1}
\hfill
F(r^0) =-g (r^0) m^0
=-\frac{\pd V(r^0)}{\pd r^0} 
=- \frac{G M m^0}{{(r^0)}^2}, 
\quad g (r^0)=\frac{G M }{{(r^0)}^2},
\quad 
V(r^0)= -\frac{GMm^0}{r^0}.
\hfill (\ref{eq-F1})
}$$

{\it \ref{Sec-microGR}.1 Effect of gravity  on the wave and particle-dynamics variables of a single $\muu$ \  } $\muu$ may be  an IED matter particle or light quantum, and is  in stationary state in a $g$ field as specified by (\ref{eq-F1}). Let firstly be no applied non-gravitational force present. At $r^0 \rar \infty$,  $g(r^0) \cdot r^0 =-V/m^0 \rar 0$  which is a maximum. Accordingly $\muu$ has an inertial mass $m^0=m(\infty)$,  mass energy $\Eng^0=m^0c^2$, and a {\it capacity to work} $\Eng^0=m^0c^2,$ which are maximum each. When brought from infinity to a finite separation $r=r(r^0)$ under $F$, $F$ has done  a negative work to $\muu$, $\D V=-\int^{r^0}_\infty F dr = - \frac{GMm^0}{r^0}=V(r^0)$ along $r$-direction. Assume that the process is (quasi) static and hence in general non-adiabatic, so no kinetic energy ($\Engvekin$) has been gained by $\muu$; $T$ would be lost,  to such as  heat. The total mechanical energy or Hamiltonian of $\muu$, hence also its {\it capacity} to do work, is thus reduced by the amount $-\D V=-V$ to 
$$\displaylines{
\refstepcounter{equation} \label{eq-H1p.a}
\hfill
H 
=  \Eng^0 + V(r^0) = m^0 c^2 - \frac{GM m^0}{r^0 }.
\hfill
 (\ref{eq-H1p.a})
}$$

$\muu$ is dually a quantum wave. For $V=0$, $\muu$ has a usual total eigen plane wave function $\psi (\rb^0,t^0) = C e^{ i(\kb^0_d \cdot \rb^0 - 2\pi \nu t^0 ) } $ (the same $\psi$
is given by the IED solution through $\psi (\rb^0,t^0) =\sum_j [\psip^j(\rb^0,t^0; \frac{\pi}{2})+i  \psip^j(\rb^0,t^0)]$) and an eigen frequency $\nu^0=m^0c^2/h$, where  $k_d^0 = (\frac{\ve}{c})k^0$, $\ve$ is particle speed and $k^0=2\pi \nu^0/c$. For a finite $V$, and $H$ as given in (\ref{eq-H1p.a}), we may establish the corresponding operator $H_{op} $, the eigen value equation and subsequently obtain (solve for)  the eigen value $H$, in a region where $V(r^0)$ is essentially constant,
$$\displaylines{
\refstepcounter{equation} \label{eq-H2p-a}
\hfill 
H_{op} \psi = H \psi, \quad  
H_{op} =  \Eng^0 + V(r^0) =m^0 c^2 - \frac{GM m^0}{r^0 } ; 
\hfill (\ref{eq-H2p-a})
\cr
\refstepcounter{equation} \label{eq-H2p} \label{eq-E1bb-b}
\hfill
H=h \nu= m^0 c^2 - \frac{GM m^0}{r^0 } = h \nu^0- \frac{GM h \nu^0 }{r^0 c^2} 
\quad {\rm or} \quad  \nu =  \nu^0(1- \frac{GM }{r^0 c^2}).
\hfill (\ref{eq-H2p})
}$$ 
Here, for $V$ is constant, the eigen function continues to be a plane wave, $\psi (\rb,t) =C  e^{ i(\kb_d \cdot \rb - 2\pi \nu t) } $. So $ i \hbar \pd \psi /\pd t = h \nu \psi$. Its equality with $H \psi$ of (\ref{eq-H2p-a}a) gives (\ref{eq-H2p}a,b). Based on (\ref{eq-H2p}), the eigen frequency $ \nu$ of the total $\muu$ wave is red shifted. If $V$ is produced by a large (spherical) mass such as the earth, all test particles in a region of  constant $r^0$ are subject to the same $V(r^0)/m$. It is thus meaningful to extend (\ref{eq-H2p}) to define here a "general-relativistic mass" $m$ of $\muu$ as
$$\displaylines{
\refstepcounter{equation} \label{eq-E0}   \label{eq-E1bb}
\hfill
 m c^2(= h \nu)= m^0 c^2 - \frac{GM m^0}{r^0 } \quad {\rm or} 
\quad  m = m^0 (1 - \frac{GM }{r^0 c^2}).
\hfill (\ref{eq-E0})
}$$

More generally, $\muu$ may be moving, at a velocity $\ve$ such that  $\g=1/\sqrt{1-\ve^2/c^2} >1$ appreciably, and subject to an applied non-gravitational potential $V_\ap$. Then $H$ is now
$$\displaylines{
\refstepcounter{equation} \label{eq-Hp}
\hfill
H' = m^0 c^2 + V  + V_\ap
= m c^2 + V_\ap,  
\quad {\rm or}
\quad 
(H' - V_\ap)^2 = m^2 c^4= m_\rest^2 c^4 + (m \ve)^2 c^2 
\hfill 
(\ref{eq-Hp})
}$$
where 
$m=\g m_\rest$, $m_\rest=\lim_{\ve^2/c^2 \rar 0} m $. The corresponding eigenvalue equation now describes the $\muu$. In typical applications, such as those in Sec \ref{Sec-exps}, $V_\ap$ is electromagnetic and varies over a quantum length scale, $a \sim 10^{-10}$ m or shorter. Across $a$, $\D V=|V(r)- V(r+a) | \dot{=}|V(r)\frac{a}{r}|$ is in general $<<|V_\ap| $\footnote{
For example, for an electron $e$ and proton $p$ at a separation $a \sim 1 \times 10^{-10} $ m,  $V_{ap}\sim -e^2/4 \pi \ev_0 a = -14.4$ eV.
For $p$ at   the sun surface, $V = -GM m_p/R= -1992 $ eV, 
$\D V \dot{=} V(R) \frac{a}{R} = 2.86 \times 10^{-16} $ eV $ << V_\ap$; 
and on the earth,  $V = -0.653$ eV,
$\D V $ is even trivially small.
};
so $V$ is essentially constant. (\ref{eq-H1p.a})-(\ref{eq-E0}) for a free $\muu$ thus hold directly.

{\it \ref{Sec-microGR}.2  Effect of gravity on the space and time co-ordinates  for the single $\muu$ \ } Based on Sec \ref{Sec-grav-pht}, gravity $F$ is transmitted to and from
 $\muu$ (at $r^0$) through the propagation of the $\psi_p$ wave (Eq \ref{eq-psig}) from and to $M$ (at $r^0=0$) at the constant speed $c$ across a distance $r^0$ given as, 
for $g \cdot r^0\rar 0$,
$$\displaylines{
\refstepcounter{equation} \label{eq-rtc}
\hfill
r^0=t^0 c
=\Ncal \tau^0 c
=\Ncal \lam^0, \quad 
\lam^0=\tau^0 c, 
\hfill (\ref{eq-rtc})
}$$ 
where $\Ncal$ is the number of wavelengths contained in $r^0$. The relationship (\ref{eq-rtc}a) firstly means that if $\psi_p$ is emitted by $M$ at $r^0=0$ at time $t=0$, it then arrives to $\muu$ at position $r^0$ after a time $t^0$. Alternatively, it also means that if the first wave front of $\psi_p$ enters $\muu$ at position $r^0$ at time $t_1^0$, it takes a further time $t^0$, i.e. at an absolute time $t_1^0+t^0$ latter, for the $\Ncal$ wavelengths to pass through $\muu$. The second meaning connects $t^0$, $r^0$ directly with the local variables $\tau^0$, $\lam^0$ at the location $r^0$, and it is the so-signified  $t^0$, and $r^0=c t^0$, that directly characterise the magnitude of $F$ in  (\ref{eq-F1}.1). In this latter sense, $r^0 =\Ncal \lam^0 $ stands for a {\it gravitational optical}, or simply "{\it gravito optical}" distance traversed by the gravity wave $\psi_p$; and $t^0$ for the time. 

For a finite $V$, the red shifted $\nu$ from $\nu^0$ in (\ref{eq-H2p}b) directly describes the particle fields $\psip$'s of $\muu$ located at $r^0$ and hence, based on (Eq \ref{eq-psig}), the $\psi_p$ wave emitted by $\muu$ here. For the $\psi_p$ wave transmitted to $\muu$ (from $M$), we can be led to the same red shifted $\nu$ by arguing simply based on Newton's law for action and reaction, assuming $M$ and $\muu$ are in dynamical equilibrium.\footnote{ 
One can formally describe the (self) effect of gravity of $M$ on its own $\psi_{p_\Msub}(r,t)$ wave, by  treating the $\psi_{p_\Msub} $ perturbed vacuuons at $r$ as a test entity. 
} 
%
So, the local wave variables for characterising the magnitude of gravity $F$ transmitted either to or from $\muu$ are the red shifted $\nu$ of (\ref{eq-H2p}b), and $\lam$ given in (\ref{eq-E2.1}a) below. Accordingly, the "gravito optical" distance $r$ and time $t$ for the $\Ncal $ number of wavelengths to pass through  (either into or out of) $\muu$ are defined by the local $\lam$ and $\tau=1/\nu$ at $r(r^0)$ given by (\ref{eq-E2.1}b,c) below:
$$\displaylines{
\refstepcounter{equation} \label{eq-E2.1}
\hfill
\lam= \frac{\lam^0}{  1 -  \frac{GM }{ r^0 c^2} },
\quad 
r = \Nscr \lam = \frac{r^0}{  1 -  \frac{GM }{ r^0 c^2}}, 
\quad  
t=\frac{r}{c}=\Nscr \tau =\frac{t^0}{  1 -  \frac{GM }{ r^0 c^2}}.
\hfill
(\ref{eq-E2.1})
}$$
(\ref{eq-E2.1}) hold irrespective of the variant $\frac{GM }{ {r'}^0 c^2}$ with $r' \approx {r'}^0 (<r)$ before $\psi_p$ arrives at $\muu$ or after $\psi_p$ has left $\muu$. Based on (\ref{eq-E2.1}b,c), it takes elongated gravito optical 
 distance and time to transmit the same $\Nscr$ wavelengths of gravity $F$. $\muu$ is thus acted by a reduced $F$, hence  able to move further apart, and re-equilibrated at the dilated  $r$. $r$ gives then the observational distance.

By virtue of the underlining second meaning of (\ref{eq-rtc}a), the effect of gravity on the space and time co-ordinates manifests exclusively along a $g$ field line. Given a $\muu$ at a distant point $(x,0,0)$ on the $x$ axis. Then in a small local  region such that every point in it is connected to $M$ by a $g$ line parallel with the $x$ axis, the general relativistic transformations from the proper co-ordinates $x^0,y^0,z^0, t^0$ to the  $x,y,z, t$ affected by the $g$ field, are given as
$$\displaylines{
\refstepcounter{equation} \label{eq-g-r-3d}
\hfill
x =\g_g x^0, \quad t= \frac{x}{c} = \g_g t^0, \quad 
\g_g= \frac{1}{1-\frac{GM}{x^0 c^2}}; 
\quad     y =y^0; 
\quad    z= z^0.
\hfill (\ref{eq-g-r-3d})
\cr
}$$
In spherical polar co-ordinates and for a point $\rb$ on  the $x$ axis, these become $r=\g_g r^0$, $t= \g_g t^0$; $\phi=\phi^0$; $\theta=\theta^0.$ Using the above for $r$, keeping $t$ as a dependent variable, the transformation for the (invariant) squared shortest line element (or geodesic) $d \sb$ of light is
$
(d \sb)^2 = -c^2 (d t)^2 + \g_g^2 [(d r^0)^2 + (r^0)^2 (d \theta^0)^2 +(r^0)^2 \sin^2 \theta^0 (d \phi^0)^2],
$
where 
$d r 
 \dot{=}   d r^0 (1-\frac{GM}{r^0 c^2})^{-1}$.

Based on experiment, Newton's law (\ref{eq-F1}) holds accurately in the $g \cdot r \rar 0$ limit. This suggests that, by retrieving from $r$  the $g \cdot r \rar 0$ value $r^0$ based on  (\ref{eq-E2.1}b), and using this in (\ref{eq-F1}), we can obtain the correct gravity, the ratios $F/m^0$ and $V/m^0$ here, as:
$$\displaylines{
\refstepcounter{equation} \label{eq-E3}
\hfill
\frac{F(r(r^0)) }{m^0(r^0)}=g(r(r^0)) =- \frac{G M }{{(r^0)}^2}  
= -\frac{ GM }{r^2\lf(1 -  \frac{GM }{ r^0 c^2}\rt)^2}, 
\quad
\frac{V (r(r^0)) }{m^0(r^0)}
= -\frac{ GM }{r \lf(1 -  \frac{GM }{ r^0 c^2}\rt)}.
\hfill (\ref{eq-E3})
}$$
From a more basic consideration, the $G$ value in Newton's law (\ref{eq-F1}) is experimentally determined on the earth (or between two Cavendish balls) which mass to $r$ ratio is small on an astronomical scale. This $G$ thus represents the $g\cdot r \rar 0 $ value. (\ref{eq-F1}) is thus expected to hold exactly if the $g\cdot r \rar 0 $ values of all other variables ($m, r$) are consistently used in it; this is done in (\ref{eq-E3}). Using (\ref{eq-E3}a) for $F$, (\ref{eq-E2.1}b) for $r^0$, the Newtonian equation of motion in the $g\cdot r\rar 0$ limit (which we are certain to be correct) is thus 
$$\displaylines{
\refstepcounter{equation} \label{eq-eom-1}
                \hfill \frac{d^2 \rb^0}{d (t^0)^2 }   \lf( =\frac{\Fb(r^0)}{ m^0}\rt) = -\frac{G M}{ (r^0)^2 } \rhat. 
\quad {\rm Or} 
\quad
\frac{d^2 \rb}{d (t^0)^2 }   
= -\frac{G M}{ r^2 (1-\frac{GM}{rc^2})^3} \rhat 
\hfill (\ref{eq-eom-1})
}$$ 
expressed using the observational $r$; the proper  $t^0$ is kept, assuming this may be 
theoretically estimated. In spherical polar co-ordinates ($r^0,\theta^0, \vphia^0$), 
$d \rb^0 = d r^0 \rhat^0 + r^0 d \theta^0 \thetahat^0 + r^0 \sin \theta^0 d \vphia^0 \vphiahat^0 $, 
$
\frac{d^2 \rb^0}{d (t^0)^2} 
=\frac{d ^2 r^0 }{d (t^0)^2 } \rhat + (\frac{d r^0}{dt^0 } + r^0\frac{d }{d t^0}) \frac{d \thetab^0}{ dt^0} 
+ (\sin \theta^0 \frac{d r^0 }{d t^0}  +r^0 \cos\theta^0 \frac{d \theta^0}{dt^0}   +r^0\sin\theta^0 \frac{d }{dt^0})        \frac{d \vphiab^0 }{d t^0}.
$ 

If an applied non-gravitational force $F_\ap$ also presents along $r^0 $ direction, similarly we can firstly write down the eom in the $g\cdot r \rar 0$ limit, $F (r^0)+ F_\ap (r^0) = m^0 \frac{d ^2 r^0}{ d (t^0)^2} $. Using the g-r transformations in it then gives the eom expressed in $r$, etc. The $g \cdot r \rar 0$ eom is a statement of the "(weak) equivalence principle". Namely, the $m^0$ acted by $F(r^0)$ or by  $F_\ap(r^0)$ is the same mass in nature; the accelerations produced by $F(r^0)$ and $F_\ap(r^0)$ are accordingly equivalent.

{\it \ref{Sec-microGR}.3 Effect of gravity on the space and time co-ordinates of
 a macroscopic test object $\mb$ \  }  $\mb$ is composed of a (large) $N$ simple single charged matter particles $\muu$'s of masses $m_\musub$'s; it has a mass $\mb=\sum_\musub m_\musub$. $\mb$ is (i) spherical, of a radius $R_\obsub$, and (ii) in internal thermal and hydrodynamic equilibrium. Within $\mb$, each $\muu,\muu'$ act on one another a central force $\Fbap_{\musub,  \musub'} $. 
 In addition, each $\muu$ is subject to the $g$ potential $V(\rb_\musub)=- \gb(\rb_\musub) \cdot \rb_\musub m_\musub$ of a large mass $M$ at $\rb=0$ at a separation  $|\rb|$ from $\mb$, where  $\rb_\musub= \rb + \langle \zetab_\musub\rangle $ and $\langle \zetab_\musub \rangle (\le R)$ are the expectation values of the distances of $\muu$ to the CM's of $M$ and $\mb$.  No other external force  presents. Assume also $\rcm >> 2 R_\obsub$ (condition iii), so $\lim_{2R/\rcm =0} \gb (\rb_\musub^0) =\gb \equiv \gb(\rb^0) = \frac{GM}{(\rcm^0) ^2} \rcmhat $, i.e. all $\muu$'s are subject to the same $\gb$. In terms of $\rb_\musub$, each $\muu$ moves according to Newton's eom
 (the correspondence principle), $m_\musub \ddot{\rb}_\musub = m_\musub \gb + \sum_{\musub' \ne \musub} \Fbap_{\musub,  \musub'}$. Sum over all $\mu$, with $\mb= \sum_\musub m_\musub$ and $\sum_\musub \sum_{\musub' \ne \musub} \Fbap_{\musub,  \musub'} =0$ under condition (ii):
$$\displaylines{
\refstepcounter{equation} \label{eq-F-ine-2}
\label{eq-F-ine-2b}
\hfill
\sum_\musub m_\musub \ddot{\rb}_\musub 
=\mb \gb + \sum_\musub \sum_{\musub' \ne \musub} \Fbap_{\musub,  \musub'}
=\mb \gb
= \mb \ddot{\rbcm}, \quad 
\rb 
= \frac{ \sum_\musub m_\musub \rb_\musub    }{\mb}
\hfill (\ref{eq-F-ine-2b})
}$$
Using (\ref{eq-E0}b), (\ref{eq-E2.1}b) for $m_\musub$, $r_\musub$ in (\ref{eq-F-ine-2b}b), with $\mb = \sum_\musub m_\musub = \sum_\musub  (1 -  \frac{GM }{ r^0 c^2}) m_\musub^0 = (1 - \frac{GM }{ r^0 c^2}) \mb^0$, $\mb^0=\sum_\musub m_\musub^0$, we obtain the scalar form of $\rb$, and time $t =r/c$,
$$\displaylines{
\refstepcounter{equation} \label{eq-F-ine-3}
\hfill 
r = \frac{| \sum_\musub m_\musub^0(1 -  \frac{GM }{ r^0 c^2}) \frac{\rb_\musub^0}{1 -  \frac{GM }{ r^0 c^2}} | }{\mb}
= \frac{\mb^0 r^0 }{ \mb^0  (1 -  \frac{GM }{ r^0 c^2}) }
=  \frac{ r^0 }{  1 -  \frac{GM }{ r^0 c^2} }, \quad t=  \frac{ t^0 }{  1 -  \frac{GM }{ r^0 c^2} },
\hfill (\ref{eq-F-ine-3})
 }$$
where $\mb^0 r^0  = |\sum_\musub m^0_\musub \rb^0_\musub| $, or $ r^0  = |\sum_\musub \frac{m^0_\musub}{\mb^0} \rb^0_\musub|$; $t^0= |\sum_\musub \frac{m^0_\musub}{\mb^0} \frac{\rb^0_\musub}{c}| $; $t^0=r^0/c$.

\section{Predictions of g-r effects in the classical test experiments of GR
} \label{Sec-exps}

We apply in this section the solutions of Sec \ref{Sec-microGR} to predict the g-r effects as manifested in the four classical-test  experiments of GR.

{\it (i) Anomalous mercury perihelion precession  \ } 
Mercury rotates along an ellipse about the sun of mass $M(=1.99 \times 10^{30}$ kg) under the sun's gravity $F$ as given by  (\ref{eq-E3}). Its eom is  (\ref{eq-eom-1}a) in terms of $r^0,t^0$. Approximating the ellipse by a circle of radius $r= a \sqrt{1-e^2} (=57.24 \times 10^{9}$ m) in the $xy$ plane here, with the sun at $r=0$, (\ref{eq-eom-1}a) becomes $- \frac{r^0 d^2 \vphia^0 }{d (t^0)^2 } \rhat = \gb^0 $; or $ -d (\frac{ d  \rb^0 }{d t^0 } ) =  \gb^0 d t^0$, where $ |d \rb^0| =r^0 d \vphia^0 (=  d (r^0 \vphia^0))$, $\frac{d \rb^0 }{d t^0} = \veb^0 $, and $d (\frac{ d  \rb^0 }{d t^0 } )= d \veb^0  = \veb^0(t^0+d t^0)-\veb^0(t^0 ) = d \veb_r^0 (\equiv d \ve_r^0 \rhat) $. Solving the equation using similar-triangles relation gives $ \frac{g^0 d t^0}{\ve^0} =\frac{|d \rb^0|}{r^0}$. Integrating over one rotation period $T^0= 2\pi r^0/ \ve^0$ gives $\frac{g^0T^0}{\ve^0}= \frac{\Phim^0 r^0}{r^0}$. Or,  $\Phim^0 = \frac{g^0 T^0}{\ve^0} = \frac{GM (T^0)^2}{ 2\pi (r^0)^3 }$ for the sweep angle in $T^0$. Using (\ref{eq-F-ine-3}.a) for $r^0$ in it gives the sweep angle ($\Phim$) in one rotation period  ${T^0}^*=2\pi r/\ve$, in terms of the observational $r$ and $\ve$ as, 
with $\Phim^0 $ computed as 
$\Phim^{0^*}=  \frac{GM ({T^0}^*)^2}{2\pi r^3 }$,
$$\displaylines{
\refstepcounter{equation} \label{eq-E4b}
\hfill
\Phim
= \frac{GM ({T^0}^*)^2}{ 2\pi r^3 (1- \frac{GM}{r^0 c^2})^3 }
\dot{=} \Phim^{0^*} (1+ \frac{3 GM }{ r c^2}). 
\quad 
{\rm Or} \quad 
\D \Phim=\Phim -\Phim^{0^*}  =  \frac{6\pi GM }{ r c^2}
  =42'' \mbox{/century},
\hfill (\ref{eq-E4b})
}$$
where $\Phim^{0^*}  =2\pi $ follows from geometric definition (in flat space) for one pure rotation in  ${T^0}^*$.
	%
$\D \Phim$ is an extra, anomalous angle swept by a fixed point, e.g. the perihelion 
(across a closed path to a precessing observer) in ${T^0}^*$, and is manifestly an angle of
precession. The observational value is $ (\D \Phim)^{exp}=43''$/century 
\cite{Exp-tests-of-GRa}.
           %
Mercury anomalous precession, and gravitational red shifts \cite{Exp-tests-of-GRb2}, were the available experiments when A Einstein devised (and justified) his GR \cite{Einstein1911a,Einstein1911b}. 


{\it (ii) Gravitational red shift \ } 
A light pulse $\muu$ emitted in the $g$ field of the sun is according to Sec \ref{Sec-microGR} red shifted, from $\lam^0$ to $\lam(R) \dot{=} \lam^0(1+ \frac{GM}{R c^2})$
 (Eq  \ref{eq-E2.1}a) for the wavelength at the sun surface $r=R(R^0)$; here  $V/m_\musub=- \frac{GM }{R^0 }$.  After further traversing a distance $r_{_{s  e}}-R \dot{=}r_{_{s e}}  (=1.496 \times 10^{11}$ m) in time $t_{se}\dot{=}\frac{r_{_{s e}}}{c}$ to reach  the earth, the instantaneous virtual source of $\muu$ (Secs \ref{Sec-grav-pht}.1-2)  is decelerated to velocity  $\ve_g = g(R)  \cdot 0-g(r_{_{s e}}) \cdot t_{_{se}} 
=\frac{GM }{r_{_{se}}^0 c}$. So formally $ \lam(R) $ is further Doppler red shifted to
$$\displaylines{
\refstepcounter{equation} \label{eq-sun-nu}
\hfill
\lam (r_{_{s e}}) 
= \lam(R) (1+\frac{\ve_g}{c}) =\lam(R) (1+ \frac{GM }{r_{_{se}}^0c^2})
  \dot{=} \lam(R) 
\dot{=} \lam^0 (1+\frac{GM }{Rc^2})       \quad (\mbox{for $r_{_{s e}} >>R$})
\hfill 
\cr
\hfill
{\rm or} \quad
\frac{\lam(r_\sesub)-\lam^0  }{\lam^0} 
(\dot{=} \frac{ g (R)\cdot R}{c^2})
 \dot{=}\frac{GM}{R c^2 }
=  2.11 \times 10^{-6}.
\hfill
(\ref{eq-sun-nu})
}$$
The predicted value agrees with the experiments \cite{Exp-tests-of-GRb}
within accepted tolerance, and the prediction of Einstein's GR \cite{Einstein1911b}.

{\it (iii) Light bending \ } Consider a light pulse $\muu$ emitted by a distant star ($s$) just  by-passing the sun (S) limb at the radius $R(R^0)$, and arriving to the earth ($e$). Let $s,S,e$ be in the $xy$ plane, the CM of the sun be at $r=0$, and the $\muu$ path $sS-Se$ be such that it crosses the $x$ axis at $x=x_0=R(R^0)$; $sS-Se$ would coincide with $x=x_0$ if under zero gravity. But the sun presents a gravity $\Fb$; for $sSe >> 2R$, $\Fb$ may be approximated as $F_x =-g m^0$ in $-x$ direction, with $g= \frac{GM}{(R^0)^2 }$ everywhere in $ R \ge y \ge -R$; and  $F_x=0$ elsewhere (the usual "thin lens approximation").  $\muu$ thus needs be emitted at an angle $\a_s$ from the vertical line $x=x_0$, such that the $\muu$ path firstly goes from $(x,y)= (R,R)$ to $(x_0,0)$, where $x_0 = R + R \tan \a_s - \D x_s=R$. This requires $R \tan \a_s -\D x_s =0$, or $\a_s\dot{=} \tan \a_s (=\frac{\D x_s}{y})=\frac{\D x_s}{R}$ (for $\frac{\D x_s}{R} <<1$). $\D x_s= \D x_{s}'+\D x_{s}'' $ is an extra $x$-component distance including two terms:
     (I) $\D x_{s}'  =  \ve T^0 = \frac{GM }{ c^2 }$ traversed by the instantaneous virtual source of $\muu$ (Sec \ref{Sec-grav-pht}), which is accelerated under $F_x$ in $-x$ direction to  velocity $\ve  =- g T^0$ in time $T^0=\frac{R^0}{c} $, and  
    (II) $\D x_{s}'' = c (T-T^0)\dot{=} \frac{GM}{ c^2}$ traversed by the radiation field comprising $\muu$ at the speed $c$, as the result of time dilation from $T^0$ to $T\dot{=}T^0 (1+ \frac{GM}{R^0c^2})$ (Eq \ref{eq-E2.1}c) under $F_x$ in $-x$ direction.
Further from $(x,y) =(R,0)$ to $(R-\D x_e,-R)$, the $\muu$ path is bent similarly, such that $\a_e\dot{=} \tan \a_e =\frac{ \D x_e }{R} $; but $\D x_e =\D x_s$ and $\a_e=\a_s$ for $F_x$ is symmetric about the $x$ axis. The total bending angle ($\a_b$) is the sum
$$\displaylines{
\refstepcounter{equation} \label{eq-alf-1a}
\hfill
\a_b= \a_s +\a_e \dot{=} \frac{\D x_s + \D x_e }{R} 
 = \frac{2  \D x_s }{R}
= \frac{2 (\D x_{s}' + \D x_{s}'') }{ R }
=\frac{4  GM }{R c^2}.
\hfill (\ref{eq-alf-1a})
}$$   
The prediction agrees with experiment \cite{Exp-tests-of-GRc}.
 Einstein's GR gives the same result \cite{Einstein1911c}.

{\it (iv) Shapiro time delay  \ } Consider a radar signal $\muu$ sent from the earth ($e$) to a planet ($p$) and reflected back during a superior conjunction, i.e. viewed from  $e$, $p$ is just above the  sun ($s$). Let $e,s,p$ be in the $xy$ plane,  $s$ be at $\rb=0$, and the $\muu$ path be just by-passing the sun limb at $x=R$. Under the sun's gravity $\Fb(\rb)= -g(\rb) m_\musub \rhat$ (Eq \ref{eq-E3}a) the distance of $\muu$ to $s$ is dilated from $r^0$ to $r\dot{=} r^0 (1+ \frac{GM}{r^0 c^2})$ (Eq \ref{eq-E2.1}b). Omitting its negligible  bending under $\Fb$ (cf Sec \ref{Sec-microGR}.iii), the geometric $\muu$ path is just the straight line $ep (pe)$ along the $y$ direction; a point $\rb (x,y)$ to $y=0$ on it 
has the $y$ co-ordinate $y= \sqrt{r^2 -R^2}$;  and $x=R$ everywhere.  A differential line element  at $\rb(R,y)$ on the $ep$ line is $d \rb \dot{=} d y \,\yhat $; $d y =\frac{dr}{\sin \phi } =  \frac{  rdr }{y } $. Or, with (\ref{eq-E2.1}b) for $r$, and $d r \dot{=} d r^0(1+ \frac{GM}{r^0 c^2})$, 
 $$\displaylines{
\refstepcounter{equation} \label{eq-dy-a}
\hfill
d y= \frac{  rdr }{y } 
=  \frac{\lf(1+ 2  \frac{GM}{r^0 c^2} +  (\frac{GM}{r^0 c^2})^2\rt) r^0 d r^0}{y};
\quad
dy-dy^0
\dot{=}\frac{r_s  d r}{ \sqrt{r^2 - R^2} } + \frac{  r_s^2  d r}{4 r \sqrt{r^2 - R^2} }, 
\hfill (\ref{eq-dy-a})
}$$ 
where $dy^0 \dot{=}  \frac{  d r^0}{ (y/r^0)} $ is to leading order the $g\cdot r\rar 0$ value of $dy$; and $ r_s = \frac{2GM}{ c^2} = 2.95 \times 10^3$ m. $y$ is not expanded for its $g$-dilation is of a higher order. $dy-dy^0 $ is the extra (gravito optical) distance travelled by $\muu$ in the $g$ field, and $d \tau = \frac{dy-dy^0}{c} $ gives the extra time. The total extra time from $R$ to $r$ in a one-way trip is
 $$\displaylines{
\refstepcounter{equation} \label{eq-dy}
\hfill
\Tcal_{1}|_{_{R \rar r}}
= \int_{t(R)}^{t(r)} d \tau
=\frac{1}{c}  \int_R^{r} \lf(\frac{r_s  }{ \sqrt{{r'}^2 - R^2} } + \frac{          r_s^2  }{4 r' \sqrt{{r'}^2 - R^2} }\rt) dr'
 = \frac{r_s}{c}  \ln \frac{r+ \sqrt{r^2-R^2} }{R} +\Ocal
\hfill (\ref{eq-dy})
}$$
where $\Ocal=\frac{r_s^2}{4Rc} \sec^{-1}\lf(\frac{r}{R}\rt)$. The total extra time, or Shapiro time delay, for $\muu$ in a round trip between the earth (at $r_e=1.496 \times 10^{11}$ m) and mars  (at $r_p=2.28 \times 10^{11}$ m) is, omitting $\Ocal$, 
 $
\Tcal_{{\rm round}}= 2( \Tcal_{1}|_{_{R \rar r_e}} + \Tcal_{1}|_{_{R \rar r_p}} )
= (2 r_s/c)  \ln [(r_e+ \sqrt{r_e^2-R^2})(r_p+ \sqrt{r_p^2-R^2} )/R^2] =247 \ {\rm \mu s}.
$
The measured value is 250 $\mu$s \cite{Exp-tests-of-GRd}.  

 The author thanks Professor C Burdik for providing the opportunity to present this  research at the ISQS25 in Prague, June, 2017, where the author has enjoyed communications with Dr Dahm, Professor S Catto, and other ISQS-25 participants. Professor B Johansson has given valuable moral support for the unification research. P-I Johansson has privately financed this research and given moral support.

\end{document}